\documentclass[pre,twocolumn,superscriptaddress,nofootinbib]{revtex4-2}
\usepackage{graphicx,amssymb,amsfonts,amsmath,chemarr,color,commath,braket,hyperref}

\begin{document}

\title{Cell sensing: from physical limits to active behaviors}

\author{Andrew Mugler}
\email{andrew.mugler@pitt.edu}
\affiliation{Department of Physics and Astronomy, University of Pittsburgh, Pittsburgh, Pennsylvania 15260, USA}

\author{Maria Rose}
\affiliation{Department of Physics and Astronomy, University of Pittsburgh, Pittsburgh, Pennsylvania 15260, USA}

\begin{abstract}
Physics sets the information contained in the signals that cells sense. But cells are active, not passive, sensors. They shape and reshape both the environment and themselves. These active behaviors allow cells to amplify, redistribute, share, and prioritize sensory information, often surpassing or obviating passive physical limits. Here, we review recent results on active sensing. After describing classic and more recent limits to sensory precision, we focus on four ways that cells implement active sensing: coordinating with other cells, reshaping their environment, dynamically updating themselves, and discriminating signals. We conclude with potential future directions.
\end{abstract}

\maketitle


\section{Introduction}

Fifty years ago, Howard Berg and Edward Purcell demonstrated that the precision of concentration sensing by bacteria is fundamentally limited by the physics of molecular diffusion \cite{berg1977physics}. Their work initiated a more general question: what can physics tell us about sensory biology? Since then, this question has been applied to a variety of sensory signals and cell types and compared to a wealth of experimental data. In many cases, cells appear to come impressively close to the precision limits set by the physics of the sensory signals.

More recently, this picture has been augmented by the recognition that cells are not passive sensors. While sensing, they reshape both their environment and themselves, they adapt and compute, and they coordinate with other cells. An emerging theme is that these active behaviors may allow cells to surpass, modify, or ignore the physical limits that would otherwise apply to passive, inanimate devices.

\begin{figure*}
\includegraphics[width=.9\textwidth]{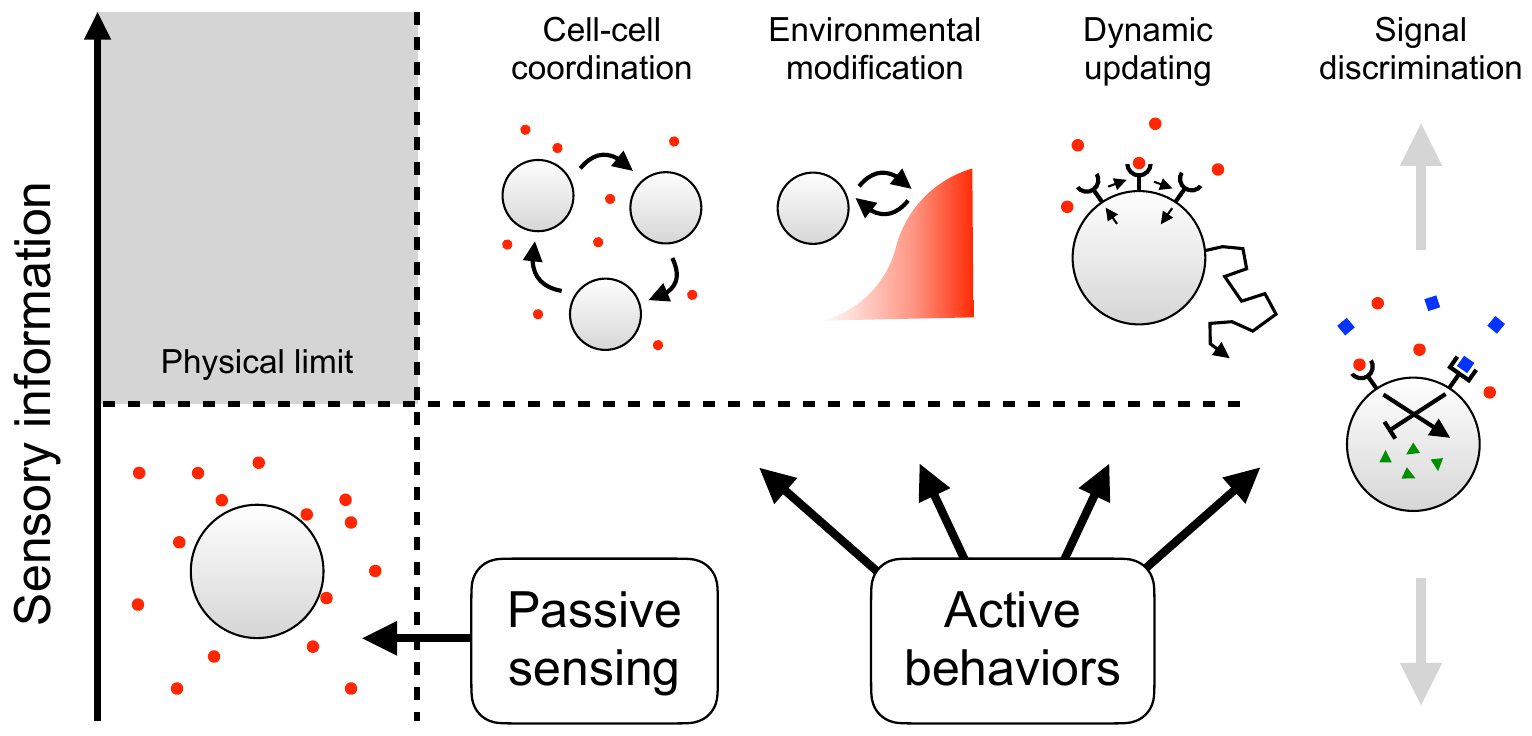}
\caption{The physics of environmental signals limits the information that can be passively sensed (left). Active behaviors, including cell-cell coordination, environmental modification, and dynamic updating of both cell and receptors, allow cells to augment or surpass these limits (middle). Internal molecular networks enable signal discrimination that can tune up or down the sensory information retained, depending on the signal type (right).}
\label{overview}
\end{figure*}

Here we review recent results on active cell sensing. We begin by describing both classic and recent results on the physical limits to sensory precision for five ubiquitous signal types. We then focus on recent progress regarding four types of active sensing ({\bf Fig.\ \ref{overview}}): (1) collective sensing via cell-cell coordination, (2) modifying the environment to better sense it, (3) dynamic updating during sensing at the cell and molecular scales, and (4) discrimination of sensory signals via networks inside the cell. We discuss how each of these active behaviors can redefine the physical constraints on the sensory process, or can change the degree to which cells approach the relevant precision limits. We conclude with key open questions and potential future directions in the field.

\section{Physical limits}

We begin by reviewing precision limits for five signal types that are ubiquitously sensed by cells and useful as baselines when we discuss active sensing. Examples we do not discuss but where precision limits have been investigated include mechanical stiffness \cite{beroz2017physical, beroz2020physical} and concentrations that change in time \cite{mora2010limits, mora2019physical} or are confined in space \cite{bicknell2015limits, gonzalez2025quantifying}.

\subsection{Concentration sensing}

Berg and Purcell considered the precision with which a cell of lengthscale $a$ could sense a molecular concentration $c$ \cite{berg1977physics} ({\bf Fig.\ \ref{limits}a}). They argued that the number of molecules near the cell at a given time scales as $n\sim ca^3$, but this number will fluctuate due to molecular diffusion. Diffusive fluctuations are Poissonian, meaning the variance equals the mean, $(\delta n)^2 \sim ca^3$. If the cell integrates many measurements over a time $t$, this variance is reduced by $t/\tau$, where $\tau$ is the signal autocorrelation time. For a diffusive signal we have $\tau \sim a^2/D$, the time needed for a molecule to diffuse with coefficient $D$ across the cell. Altogether, the fractional error in concentration estimation $\delta c/c = \delta n/n$ scales as
\begin{equation}
\label{conc}
\frac{\delta c}{c} \sim \frac{1}{\sqrt{caDt}}.
\end{equation}
Berg and Purcell supported their arguments with a rigorous derivation of this expression and extended it to the case of one or more receptors detecting the concentration rather than the whole cell \cite{berg1977physics}.

They went on to show that an expression like Eq.\ \ref{conc} predicts a minimum time of approximately one second that a bacterium must swim before changing directions if it is to detect with sufficient precision whether the concentration is increasing or decreasing. The fact that bacteria actually do swim for about one second led them to conclude that bacterial sensory machinery closely approaches the physical limit of Eq.\ \ref{conc}. Interestingly, experiments in the past few years have brought fresh scrutiny to this conclusion \cite{mattingly2021escherichia, mattingly2026coli}, a point to which we will return when discussing future directions.

In the years since Berg and Purcell's work, researchers have revisited the problem of concentration sensing to account for the kinetics, cooperativity, and spatiotemporal correlations involved in receptor binding \cite{bialek2005physical, bialek2008cooperativity, endres2009maximum, skoge2011dynamics, skoge2013chemical, kaizu2014berg}. These cell-intrinsic effects can increase the error above that in Eq.\ \ref{conc}, which only accounts for the extrinsic noise in the signal itself. In this sense, Eq.\ \ref{conc} sets a physical noise floor \cite{bialek2005physical}.

\subsection{Gradient sensing}

\begin{figure*}
\includegraphics[width=\textwidth]{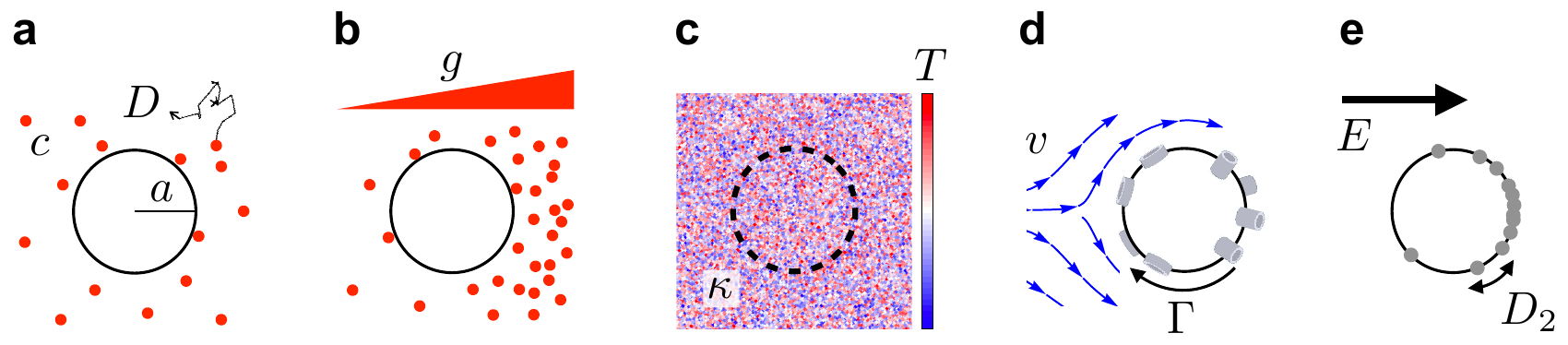}
\caption{Physical limits to sensing precision. {\bf (a)} Concentration sensing: a cell with lengthscale $a$ senses molecules with concentration $c$ and diffusion coefficient $D$. {\bf (b)} Gradient sensing: a cell senses $g$, the change in concentration per length. {\bf (c)} Temperature sensing: the temperature $T$ of fluid with thermal conductivity $\kappa$ fluctuates throughout the cell volume. {\bf (d)} Fluid flow sensing: Flow with speed $v$ induces a tension difference $\Gamma$ in the cell membrane, opening more mechanosensitive ion channels on the upstream side. {\bf (e)} Electric field sensing: Field with strength $E$ biases charged sensor molecules that diffuse in the membrane with coefficient $D_2$.}
\label{limits}
\end{figure*}

Later work considered the precision with which a cell could sense a concentration gradient $g$ (concentration per length) \cite{endres2008accuracy, endres2009accuracy, hu2010physical} ({\bf Fig.\ \ref{limits}b}). The argument is similar to that above, but now the cell estimates a molecule number difference between its two halves, $\Delta\sim(ga)a^3$ \cite{mugler2016limits}. The variance is still set by the background concentration, $(\delta \Delta)^2 \sim ca^3$, and the autocorrelation time is still set by diffusion, $\tau\sim a^2/D$. Therefore, $\delta g/g=\delta\Delta/\Delta$ scales as
\begin{equation}
\label{grad}
\frac{\delta g}{g} \sim \frac{1}{ga}\sqrt{\frac{c}{aDt}}.
\end{equation}
Notably, the error increases with $c$ because it is harder to measure a small difference on top of a large background \cite{mugler2016limits}.

Endres and Wingreen \cite{endres2008accuracy} derived an expression like Eq.\ \ref{grad} more rigorously and compared it to data \cite{van2007biased} on amoeba migrating up a gradient of a chemical to which they are attracted, a process called chemotaxis. They found that cells would need to integrate their measurements for at least three seconds to explain the observed chemotactic response. The fact that this timescale is commensurate with amoebae's actual response times again suggests that these cells closely approach the physical limit of Eq.\ \ref{grad} \cite{varennes2016sense}.

\subsection{Temperature sensing}

The precision with which a cell could sense temperature was considered forty years ago by Dusenbery \cite{dusenbery1988limits}. More recently, this work was extended to derive a physical limit \cite{vennettilli2021precision} ({\bf Fig.\ \ref{limits}c}). The basic argument is that the variation in the temperature $T$ of a volume of fluid scales as $(\delta T/T)^2 \sim k_{\rm B}/C$, where $C$ is the heat capacity and $k_{\rm B}$ is Boltzmann's constant. Taking a volume the size of a cell gives $C\sim c_s\rho_w a^3$, where $c_s$ and $\rho_w$ are the specific heat and density of the fluid, respectively. Integrating over a time $t$ reduces this variance by $t/\tau$ as above, where now $\tau\sim a^2/D_T$ depends on the thermal diffusivity $D_T = \kappa/(c_s\rho_w)$, with $\kappa$ the thermal conductivity. Putting things together, $c_s$ and $\rho_w$ drop out, and we get
\begin{equation}
\label{temp}
\frac{\delta T}{T} \sim \sqrt{\frac{k_{\rm B}}{\kappa at}}.
\end{equation}
Using the thermal conductivity of water and a cell size on the order of microns, this expression predicts that temperature can be estimated to one part in a million in less than a microsecond \cite{vennettilli2021precision}. Clearly this is a case where the physical limit is not very limiting: the noise floor for temperature sensing is extremely low.

Instead, the precision of temperature sensing is limited by a cell's intrinsic components. Some bacteria sense temperature using protein dimerization and achieve an estimation error on the order of $10\%$ \cite{hurme1997proteinaceous, vennettilli2021precision}. In this case the precision is ultimately limited by molecule counting noise like in the concentration sensing examples above \cite{vennettilli2021precision}. More strikingly, neurons in the thermal imaging organs of snakes can respond to milli-Kelvin temperature changes \cite{bakken2007imaging}, despite the fact that the ion channels that perform the sensing are only sensitive to single-Kelvin changes \cite{gracheva2010molecular}. Here the thousand-fold sensitivity increase is thought to arise from integrating many such sensors in the vicinity of a dynamical bifurcation that preserves sensory information in the presence of noise \cite{graf2024bifurcation}.

\subsection{Fluid flow sensing}
\label{Sflow}

Many cells respond to fluid flow, e.g., by migrating upstream or downstream. Here, like in temperature sensing, variation in the properties of the fluid itself is very small, and precision is instead set by the cell-intrinsic sensory mechanism. A precision limit has been derived for the mechanism of sensing flow direction using mechanosensitive ion channels \cite{bouffanais2013physical} ({\bf Fig.\ \ref{limits}d}), and the argument is as follows \cite{gonzalez2025quantifying}. Membrane tension is higher on the upstream side of the cell, increasing the probability $p$ that a channel there is open. For a two-state switch in a thermal bath, statistical mechanics tells us $p = 1/(1+e^{-W/k_{\rm B}T})$, where $W$ is the energy difference between the closed and open states. The energy difference is $W = \gamma S$, where $\gamma$ is the added tension near that channel due to the flow, and $S$ is the surface area difference between the open and closed states \cite{ursell2008role}. Generally $\gamma S\ll k_{\rm B}T$, such that $p \approx 1/2 + \gamma S/(4k_{\rm B}T)$. For $M$ independent channels, the difference in the number of open channels between the two halves of the cell therefore scales as $\Delta \sim M\Gamma S/(k_{\rm B}T)$, where $\Gamma$ is the tension difference across the cell. The variance in $\Delta$ is the sum of binomial variances for each switch, $(\delta\Delta)^2\sim Mp(1-p) \approx M/4$. Once again, this variance is reduced by $t/\tau$, where here the autocorrelation time $\tau \sim k^{-1}$ is set by the switching rate $k$. Altogether,
\begin{equation}
\label{flow}
\frac{\delta \Delta}{\Delta} \sim \frac{k_{\rm B}T}{\Gamma S\sqrt{Mkt}}.
\end{equation}

The relationship between the fluid flow and the tension difference $\Gamma$ depends on the permeability $K$ of the environment. Here $\sqrt{K}$ is the characteristic distance between obstacles through which the fluid flows, to be compared with the cell lengthscale $a$. In purely aqueous environments ($\sqrt{K} \gg a$), a flow with speed $v$ induces a shear near a surface with rate $G \sim v/a$, and the tension difference scales as $\Gamma \sim G\eta a \sim v\eta$ \cite{bouffanais2013physical}, where $\eta$ is the fluid viscosity. In contrast, in extracellular matrix ($\sqrt{K} \ll a$), the flow induces a pressure difference $P \sim v\eta a/K$ \cite{polacheck2014mechanotransduction}, and the tension difference scales as $\Gamma \sim Pa \sim v\eta a^2/K$ via the Young-Laplace equation. We see that low permeability boosts the tension difference by a factor $a^2/K\gg 1$, in principle allowing the detection of much lower flow speeds.

Indeed, in aqueous environments, eukaryotic cells including endothelial cells \cite{olesen1988haemodynamic}, amoebae \cite{decave2003shear}, neutrophils \cite{makino2006g}, and fibroblasts and stem cells \cite{park2010cell} respond to shear stresses in the range $G\eta \sim 0.01$$-$$1$ Pa, corresponding to flow speeds in the range $v\sim Ga \sim 0.1$$-$$10$ mm/s. In matrix environments, cancer cells respond to much slower flow speeds of $v \sim 1$ $\mu$m/s \cite{polacheck2011interstitial, polacheck2014mechanotransduction}. The physical reason is that the low permeability of the matrix ($K \approx 0.1$ $\mu$m$^2$ \cite{polacheck2011interstitial}) leads to pressure differences $P \sim v\eta a/K\sim0.1$ Pa in the same range as the shear stress.

Whether cells achieve the limit of Eq.\ \ref{flow} will be addressed later when we discuss signal discrimination.

\subsection{Electric field sensing}
\label{Selectric}

Cells are thought to sense an electric field using charged molecules on the cell membrane that are redistributed by the field \cite{mclaughlin1981role, brown1994electric, allen2013electrophoresis, kobylkevich2018reversing, sarkar2019electromigration}. A physical limit for this process has been derived \cite{nwogbaga2023physical}, and we present a basic argument here.
The key idea is that the sensor molecules are pushed by an electric field toward one side of the cell, but they also diffuse in the membrane which promotes uniformity ({\bf Fig.\ \ref{limits}e}). The tradeoff results in an angular distribution $p(\theta)$ whose width decreases with the dimensionless parameter $\epsilon = a\mu E/D_2$, where $D_2$ is the diffusion coefficient, $\mu$ is the sensors' electric mobility, and $E$ is the electric field strength \cite{nwogbaga2023physical}. This parameter is a P\'eclet number, comparing advection, at velocity $\mu E$, to diffusion, over the lengthscale of the cell. An estimate from $M$ independent sensors obeying $p(\theta)$ has a variance $(\delta\theta)^2 \sim 1/(M\epsilon)$ for $\epsilon\gg1$ or $(\delta\theta)^2 \sim 1/(M\epsilon^2)$ for $\epsilon\ll1$ \cite{mardia2009directional, nwogbaga2023physical}, and experiments suggest the latter \cite{nwogbaga2023physical}.  As usual, the variance can be reduced by $t/\tau$, where here $\tau \sim a^2/D_2$ is set by the time for a sensor to diffuse across the cell membrane. Altogether,
\begin{equation}
\label{electric}
\delta\theta \sim \frac{1}{\mu E}\sqrt{\frac{D_2}{Mt}}.
\end{equation}
This expression is valid for $\delta\theta \ll 1$ and otherwise acquires corrections due to periodicity on a circle \cite{nwogbaga2023physical}. Because membrane diffusion is slow, $\tau$ is large, and cells may not time-average significantly before responding, in which case Eq.\ \ref{electric} would lack the division by $t/\tau$. With these considerations, a result like Eq.\ \ref{electric} compared favorably to electric field-induced migration data on three cell types and implied plausible values for the underlying parameters \cite{nwogbaga2023physical}.

\section{Cell-cell coordination}

How do the above limits change when sensing is active? We now review work on active sensing mechanisms that can address this question. We organize our discussion ``from the outside in,'' first discussing multicellular effects, then focusing on a cell's local environment, then describing dynamic updating at the cell and molecular levels, and finally considering the role of signal processing inside the cell.

\begin{figure*}
\includegraphics[width=\textwidth]{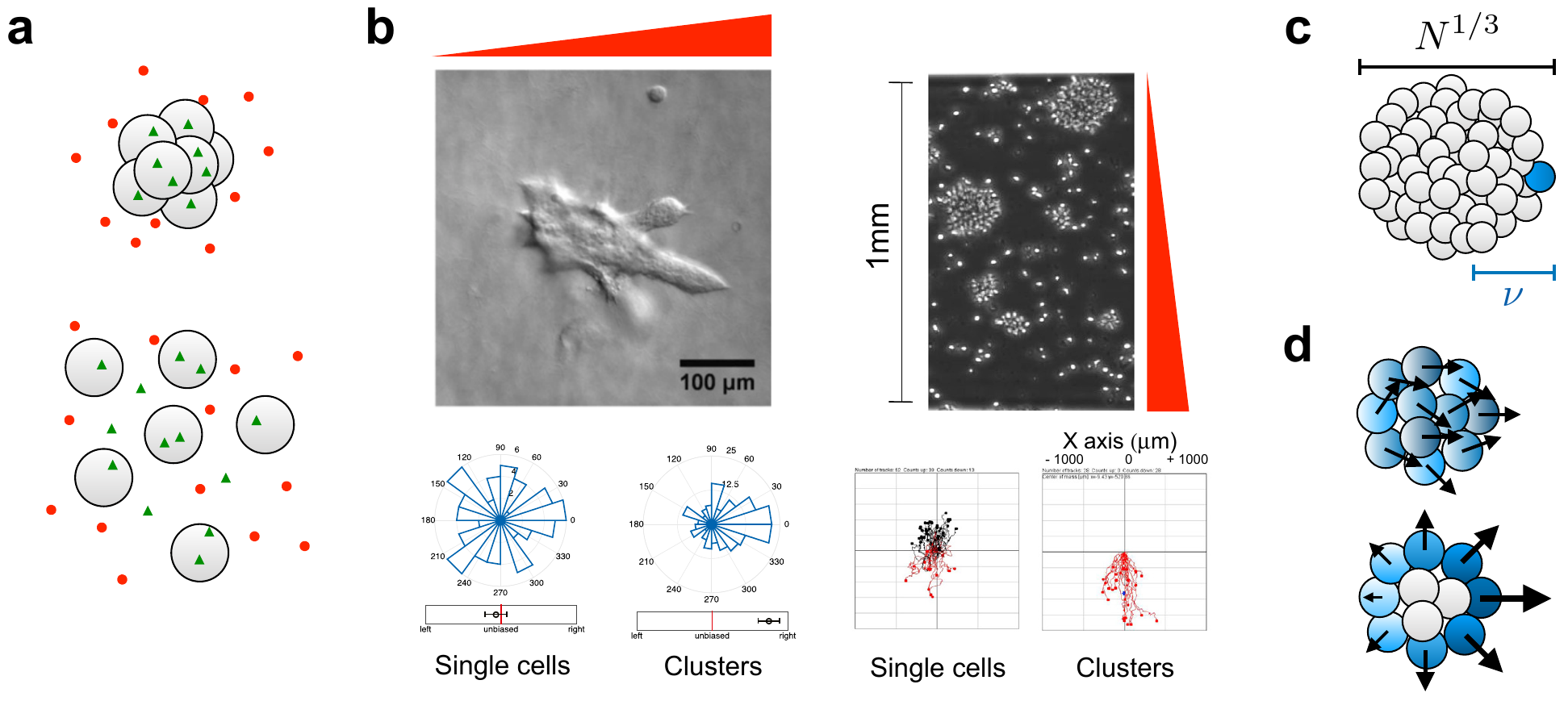}
\caption{Active sensing by cell-cell coordination. {\bf (a)} Short-range (top) or long-range (bottom) chemical communication (green) improves the precision with which each cell senses a concentration (red). {\bf (b)} Clusters of epithelial cells (left) and immune cells (right) respond to gradients to which their individual cell constituents do not respond alone. Adapted from \cite{ellison2016cell} and \cite{malet2015collective}, respectively. {\bf (c)} Due to imperfect communication, a responding cell (blue) receives information from only a subset $\nu$ of the $N^{1/3}$ cells along a gradient direction. {\bf (d)} Collective chemotaxis can be individual-driven (top) or emergent (bottom).}
\label{collective}
\end{figure*}

\subsection{Collective concentration sensing}

Chemical cell-cell communication is widespread. In principle, a cell could improve the precision of a concentration estimate using information received from other cells. Fancher and Mugler derived the precision limit for such a strategy, considering both short-range (cell-to-cell) and long-range (diffusive) communication \cite{fancher2017fundamental} ({\bf Fig.\ \ref{collective}a}). They found
\begin{equation}
\label{concN}
\frac{\delta c}{c} \sim \frac{1}{\sqrt{cN^\alpha aDt}},
\end{equation}
where $N$ is the number of cells in the population, and $\alpha$ is a scaling exponent. Comparing with Eq.\ \ref{conc}, we see that the factor of $N^\alpha$ reduces the estimation error below that for an isolated cell. Moreover, they found that for short-range communication, $\alpha = 1/3$ because cells act as a spherical ``supercell'' with a radius that scales as $N^{1/3}a$. In contrast, for long-range communication, $\alpha > 1/3$, improving the precision further, because cells can spread out to make concentration measurements that are more statistically independent. An optimal cell-cell separation results from the tradeoff between spreading out to reduce estimation error and staying close to maintain communication strength \cite{fancher2017fundamental}. This separation effect agrees with observations of brain tumor cells that exhibit long-range communication and autonomously adopt separations of several cell diameters \cite{kravchenko2014glioblastoma}. More generally, cells are thought to exploit population-averaging of diffusible signals in contexts where precision is paramount, such as boundary formation in developing embryos \cite{gregor2007probing, erdmann2009role, fancher2020diffusion}.

\subsection{Collective gradient sensing}

Strikingly, experiments have demonstrated that cell collectives, including neurons \cite{rosoff2004new}, neural crest cells \cite{theveneau2010collective}, epithelial cells \cite{ellison2016cell}, and immune cells \cite{malet2015collective}, respond to concentration gradients to which their single-cell constituents do not respond alone ({\bf Fig.\ \ref{collective}b}). As above, a cell leading such a collective response could improve the precision of its gradient estimate using information received from other cells. The estimation error of such a strategy was derived to explain the epithelial cell case \cite{ellison2016cell, mugler2016limits} and reads
\begin{equation}
\label{gradN}
\frac{\delta g}{g} \sim \frac{1}{\nu ga}\sqrt{\frac{c}{aDt}},
\end{equation}
where $\nu > 1$ is the effective number of cells with which the leading cell communicates along the gradient direction. Comparing with Eq.\ \ref{grad}, we see that the factor of $\nu$ reduces the estimation error below that for a single cell. For perfect communication, $\nu\sim N^{1/3}$, and the precision increases indefinitely with the linear size of the collective. For imperfect communication, $\nu$ is limited by the communication rate, and the precision saturates even as the collective continues to grow \cite{ellison2016cell, mugler2016limits} ({\bf Fig.\ \ref{collective}c}). The response of the epithelial cells was observed to saturate as a function of the collective size, from which it was determined that communication effectively spans $\nu\approx 3$$-$$4$ cells \cite{ellison2016cell}.

If cell-cell coordination is mechanical rather than chemical, precision may not be limited by a smaller lengthscale within the collective. For example, in many cases \cite{theveneau2010collective, cai2016modeling, haeger2015collective}, including for the immune cells \cite{malet2015collective}, a gradient causes cells to migrate while remaining mechanically adhered. Such collective chemotaxis can occur due to one of two mechanisms: (1) each cell could sense and respond to the gradient individually but not break the adhesion; or (2) each cell could sense the local concentration, and chemotaxis could emerge from the cells' unequal responses ({\bf Fig.\ \ref{collective}d}). Both mechanisms have been investigated theoretically \cite{malet2015collective, varennes2016collective, camley2016emergent, camley2016collective, cai2016modeling, varennes2017emergent, camley2017cell, camley2018collective}, and a key result is a precision limit that has the form of Eq.\ \ref{gradN} \cite{varennes2017emergent}. In the individual chemotaxis case, $\nu\sim N^{1/2}$, consistent with error reduction by many individual measurements (in this case, cells). In the emergent chemotaxis case, $\nu$ depends on the dimensionality of the collective: $\nu\sim N^{1/2}$ for three-dimensional clusters, equivalent to the individual chemotaxis case; whereas $\nu\sim N^{3/4}$ for two-dimensional cell sheets and $\nu\sim N$ for one-dimensional cell chains, both outperforming the individual chemotaxis case \cite{varennes2017emergent}. Evidence of emergent chemotaxis exists for neural crest cells \cite{theveneau2010collective}, immune cells \cite{malet2015collective}, and ovary cells \cite{cai2016modeling}, where edge cells' outward polarization is biased by the gradient.

Collective chemotaxis can also occur in clusters that are not mechanically adhered but are colocalized by the secretion of a coattractant.
Emergent chemotaxis can still operate \cite{camley2016collective, camley2017physical}, and indeed, coattraction is known to occur in neural crest cells \cite{carmona2011complement}. Evidence for coattraction also exists in bacteria \cite{budrene1991complex, long2017cell} and has been shown theoretically to increase the speed of collective chemotaxis \cite{saha2026effects}.

\section{Environmental modification}

Many cells modify the environmental signals that they sense. A classic example occurs in bacterial chemotaxis, where cells consume a nutrient, the consumption creates a nutrient gradient, and the cells move up the gradient \cite{adler1966chemotaxis, keller1971traveling, saragosti2011directional, fu2018spatial, martinez2022morphological}. Similar effects have been observed in other systems \cite{tweedy2016self}, including epithelial cell migration \cite{scherber2012epithelial}, immune cell migration \cite{uccar2025self}, cancer metastasis \cite{scherber2012epithelial, muinonen2014melanoma}, and embryogenesis \cite{dona2013directional}. In these cases, the cells create a signal (the gradient) where none previously existed. Do cells modify an already present signal to make it easier to sense? Here we review recent progress on two such cases: degrading a gradient to enhance directional sensing, and secreting a chemical to detect a fluid flow.

\subsection{Gradient enhancement}
\label{Syeast}

\begin{figure*}
\includegraphics[width=.85\textwidth]{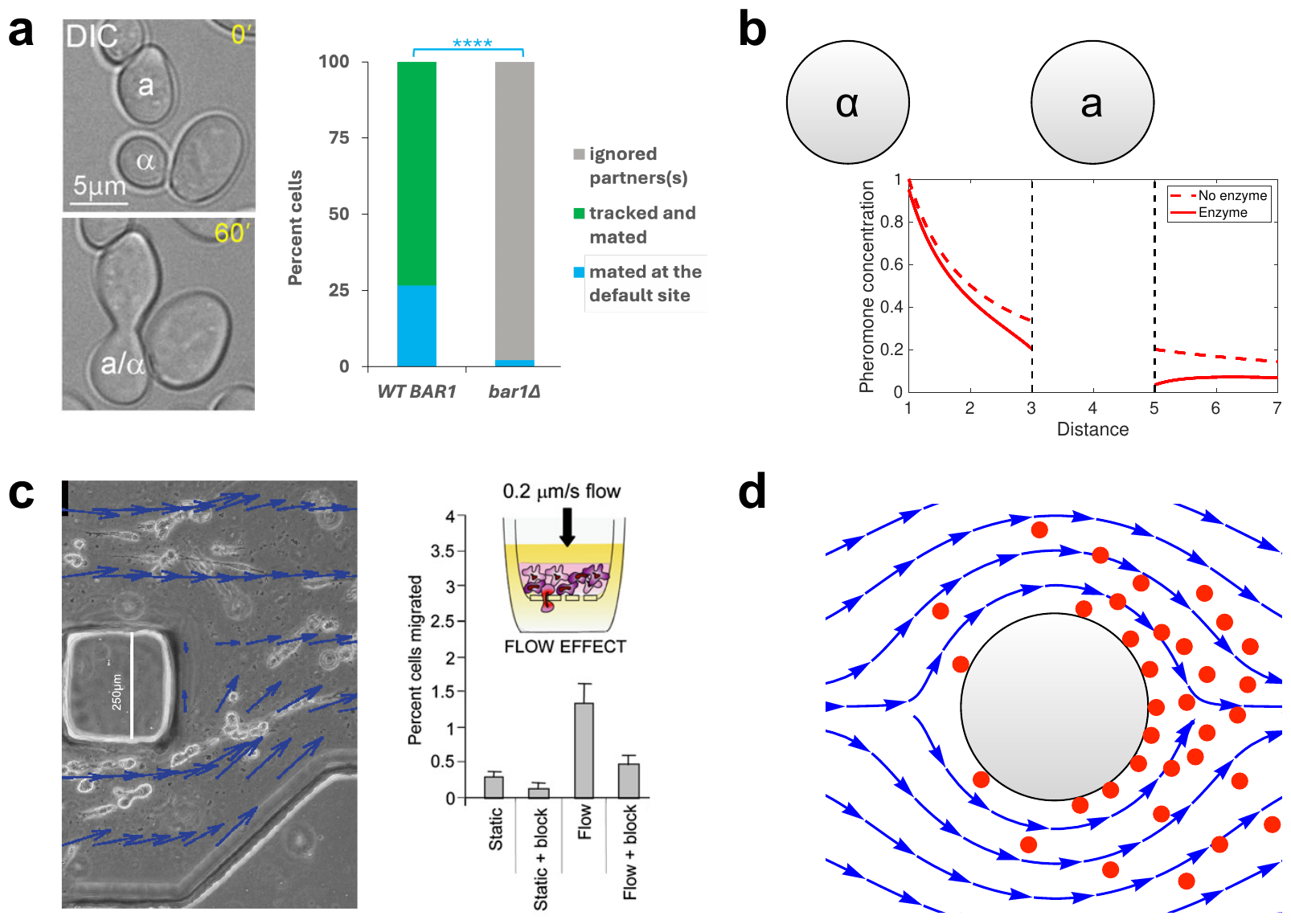}
\caption{Active sensing by environmental modification. {\bf (a)} Left: An a-type yeast cell mates with an $\alpha$-type yeast cell. Adapted from \cite{wang2019mating}. Right: Mating fails if a-cells do not secrete bar1, an enzyme that degrades the $\alpha$-cells' pheromone. Paul Urban and David Stone, unpublished data. {\bf (b)} The enzyme decreases the pheromone concentration and increases its gradient. Model from \cite{lefebre2024role}. {\bf (c)} Left: Cancer cells follow fluid flow downstream. Adapted from \cite{polacheck2011interstitial}. Right: Migration is reduced by blocking a receptor for a chemical the cells themselves secrete. Adapted from \cite{shields2007autologous}. {\bf (d)} The secreted chemical is biased by the flow, creating a gradient that the cell detects, a process called autologous chemotaxis.}
\label{environment}
\end{figure*}

During gradient sensing, cells have been observed to actively remove the chemical they are sensing. An example of short-range removal is receptor endocytosis. Upon binding the chemical, a receptor is internalized and recycled, removing chemical from the local environment \cite{irannejad2015effects}. Experiments have shown that blocking endocytosis disrupts chemotaxis in cancer cells \cite{mutch2014polarised, pinilla2023cbl} and ovary cells \cite{assaker2010spatial}. An example of long-range removal is the secretion of a degrading enzyme, which has been observed during gradient sensing in amoebae \cite{garcia2009group} and yeast \cite{jin2011yeast}. For yeast, the cell finds a mate by detecting a secreted pheromone, but it also secretes an enzyme to degrade that pheromone. The degradation is essential for efficient mating \cite{chan1982physiological, jackson1990courtship, jin2011yeast, jacobs2022mechanism} ({\bf Fig.\ \ref{environment}a}).

Why would a cell destroy the signal it aims to detect?
In the case of yeast mating, it has been proposed that the secretion of a degrading enzyme disambiguates potential mate locations \cite{jackson1990courtship, barkai1998protease}, prevents receptor saturation \cite{jacobs2022mechanism, jin2011yeast}, and sharpens the gradient of the pheromone profile \cite{andrews2010detailed, jin2011yeast, lakhani2017testing}. Endocytosis is also predicted to sharpen a gradient and increase the anisotropy, the front-back difference in the number of detected molecules \cite{barrios2026endocytosis}.

Yet, the strategy of sharpening a gradient by degrading it raises a fundamental question about sensing: how can one obtain more information by detecting fewer events? This question was addressed with a model of yeast mating \cite{lefebre2024role} ({\bf Fig.\ \ref{environment}b}), and a key result is a precision limit for the anisotropy $A$ when degradation is weak,
\begin{equation}
\label{degrade}
\frac{\delta A}{A} \sim \frac{1}{ga}\sqrt{\frac{c}{aDt(1+\beta)}},
\end{equation}
where
\begin{equation}
\label{degrade2}
c = \frac{s}{4\pi Dr_0}(1-u_1\beta), \qquad
g = \frac{s}{4\pi Dr_0^2}(1+u_2\beta).
\end{equation}
Here, $r_0$ is the separation between two mating cells, $s$ is the pheromone secretion rate, $\beta \ll 1$ is a dimensionless parameter characterizing the strength of the degrading enzyme, and $u_1$ and $u_2$ are positive factors of order unity.

Comparing to Eq.\ \ref{grad}, we see that, without the degrading enzyme ($\beta=0$), Eq.\ \ref{degrade} has the expected form for gradient sensing, with $c = s/(4\pi Dr_0)$ the concentration a distance $r_0$ away from the pheromone source, and $g = |dc/dr|_{r_0}$ its gradient there. The degrading enzyme ($\beta > 0$) has three effects: (1) it decreases the concentration, as expected; (2) it increases the gradient, as expected; and (3) it decreases the autocorrelation time due to faster pheromone turnover, leading to the factor $1+\beta$ in Eq.\ \ref{degrade}. All three effects decrease the estimation error, Eq.\ \ref{degrade}, showing that weak degradation benefits gradient sensing. Yet, sufficiently strong degradation will destroy the pheromone signal completely, preventing sensing. This tradeoff leads to an optimal degradation strength, and data suggest that yeast operate below this optimum, in the regime where degradation benefits sensing \cite{lefebre2024role}.

This example shows that it is possible to obtain more information from fewer events as long as the information per event sufficiently increases. In this case the information per event can increase only because the pheromone profile formed by diffusion and degradation is out of equilibrium \cite{lefebre2024role}. The formation of nonequilibrium profiles by diffusion and degradation is ubiquitous, notably in morphogenesis \cite{wartlick2009morphogen, gregor2007stability}, suggesting that the enhancement of gradient sensing by degradation is a generic phenomenon.

\subsection{Autologous chemotaxis}
\label{Sautologous}

Cancer cells migrate downstream in response to fluid flow \cite{shields2007autologous, polacheck2011interstitial, munson2013interstitial}, which may help them find vessels in the body during metastasis \cite{shields2007autologous}. Interestingly, migration is reduced by blocking a receptor for a chemical that the cells themselves secrete \cite{shields2007autologous, polacheck2011interstitial} ({\bf Fig.\ \ref{environment}c}). This suggests that flow sensing is aided by the secretion and detection of a chemical. The chemical is biased by the flow, resulting in more detection events on the downstream side of the cell, a process called autologous chemotaxis \cite{shields2007autologous} ({\bf Fig.\ \ref{environment}d}).

Autologous chemotaxis can be modeled using fluid dynamics and stochastic processes \cite{fleury2006autologous, fancher2020precision, khair2021two, vennettilli2022autologous, ben2024using, gonzalez2023collective, gonzalez2025quantifying}, and a precision limit for the anisotropy has been derived \cite{gonzalez2025quantifying},
\begin{equation}
\label{autologous}
\frac{\delta A}{A} \sim \frac{1}{ga}\sqrt{\frac{c}{aDt}},
\end{equation}
where
\begin{equation}
\label{autologous2}
c = \frac{s}{aD} + \frac{s\rho L}{v}, \qquad
g = \frac{s v}{aD^2}.
\end{equation}
Here, $v$ is the flow speed, $s$ is the secretion rate, $\rho$ is the cell density, and $L$ characterizes the length of the environment in the direction of the flow (e.g., a flow chamber in experiments). This result is valid for small P\'eclet number $\epsilon = va/D$, which compares the effect of flow to diffusion on the secreted molecules, over the lengthscale of the cell. The P\'eclet number is indeed small in the experiments and in the cancer microenvironment due to the low speed of interstitial flow \cite{chary1989direct, fleury2006autologous, fancher2020precision}.

We see that Eq.\ \ref{autologous} has the same form as Eq.\ \ref{grad}, where now the concentration $c$ is supplied by the cell, and the gradient $g$ is created by the flow (Eq.\ \ref{autologous2}). Specifically, the first term of $c$ comes from molecule secretion by the cell ($s$), and $g$ reflects the difference in this concentration across the cell due to the flow ($v$). The second term in $c$ is a background concentration due to secretion from other cells ($\rho$). A large background obscures gradient detection, and correspondingly this term increases the estimation error. It decreases as molecules leave the system, either due to a shorter system lengthscale ($L$) or faster flow ($v$).

For an isolated cell ($\rho=0$), Eq.\ \ref{autologous} reduces to $\delta A/A\sim 1/(\epsilon\sqrt{s t})$. With measured values of the P\'eclet number $\epsilon$ and the secretion rate $s$, this expression predicts that an estimation error of $\delta A/A \approx 30\%$ requires an integration time of at least $t\approx 15$ hours \cite{fancher2020precision}. Fifteen hours is the timeframe over which cells are observed to migrate \cite{shields2007autologous}, suggesting that cancer cells operate remarkably close to the physical limit of autologous chemotaxis.

Proximity to the physical limit implies that if the cell density $\rho$ is sufficiently increased, each cell's estimation error will become prohibitive, and autologous chemotaxis should fail. Indeed, Eq.\ \ref{autologous2} identifies a critical density $\rho_*\sim v/(aDL)$ at which the two terms in $c$ are on the same order. Measured values place this density at $\rho_*\approx 50$ cells/mm$^3$ \cite{vennettilli2022autologous}, and experiments in a flow chamber showed that cells at this density successfully migrated downstream, whereas cells at a density five times higher did not \cite{polacheck2011interstitial}.

Intriguingly, the cells at the higher density did not stop or migrate randomly; they migrated upstream. Experiments identified this behavior with a second mechanism that competes with autologous chemotaxis, sensing the flow mechanically \cite{polacheck2011interstitial, polacheck2014mechanotransduction} as discussed in Section \ref{Sflow}. This observation casts flow sensing by cancer cells as a negotiation between conflicting cues, to which we will return when discussing signal discrimination.

\section{Dynamic updating}

Thus far, most of the works we have reviewed assume that cells sit still, integrate their measurements, then act. Yet, cells are alive and dynamic. Their actions and their data acquisition likely happen at the same time and feed back on one another. Here we review progress in this realm of dynamic sensing.

\subsection{Feedback from movement}

\begin{figure*}
\includegraphics[width=.95\textwidth]{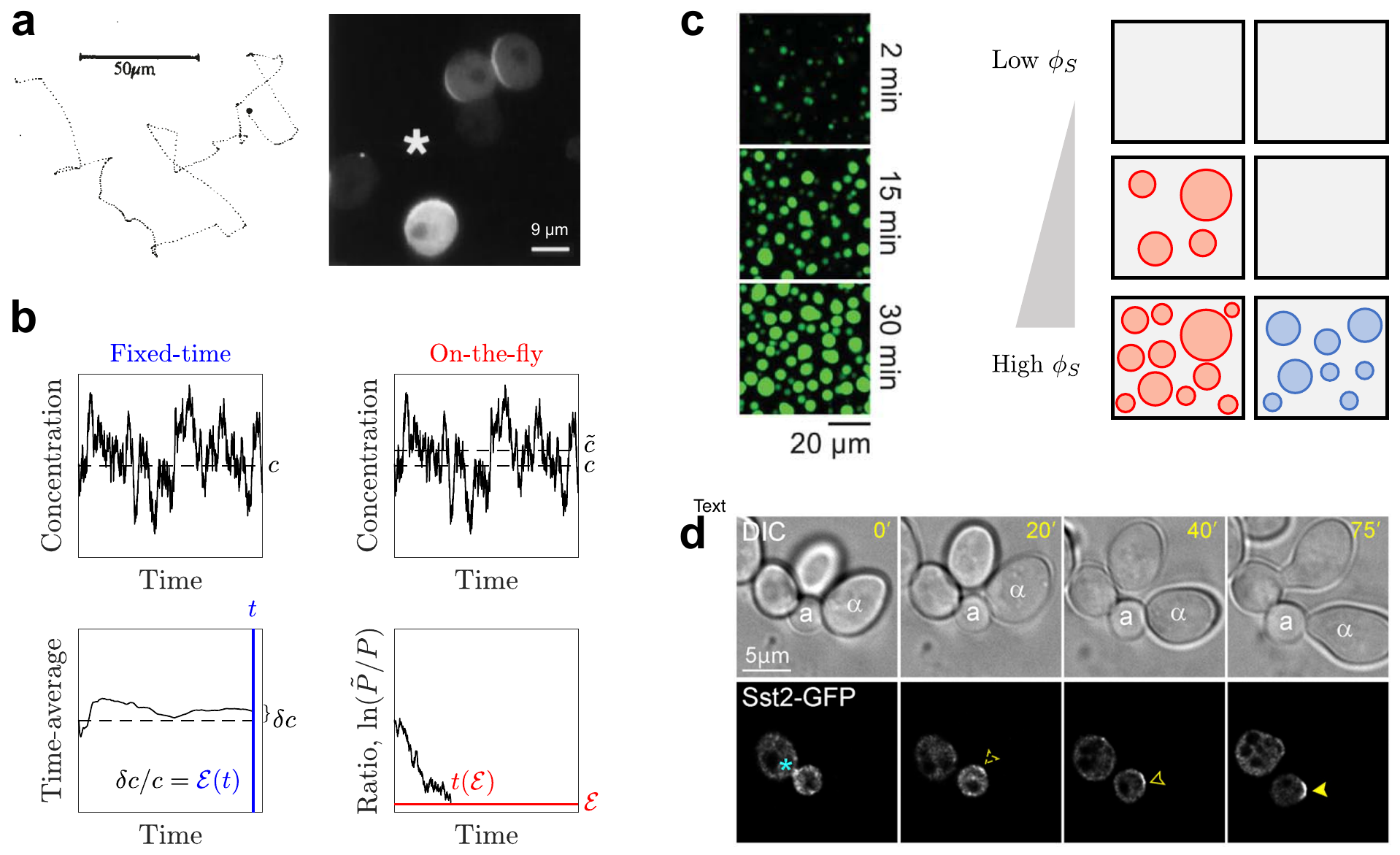}
\caption{Active sensing by dynamic updating. {\bf (a)} Sensing with and without movement. Left: A chemotactic bacterium moves in three dimensions. Dots are two-dimensional projection every $0.08$ s. Adapted from \cite{berg1972chemotaxis}. Right: Amoebae sense a chemical on one side of their cell body (bright arcs) even when prevented from moving. Asterisk is chemical's source. Adapted from \cite{parent1999cell}. {\bf (b)} Fixed-time vs.\ on-the-fly sensing. Left: Averaging a concentration $c$ for a fixed time $t$ to obtain an estimation error ${\cal E}(t)$. Right: Measuring until an alternative hypothesis $\tilde{c}$ has a sufficiently small log likelihood ratio ${\cal E}$, for a time $t({\cal E})$. {\bf (c)} Sensing by phase separation. Left: a molecular component separates into droplets in the presence of a second component, both in vitro (shown) and in vivo. Adapted from \cite{du2018dna}. Right: Schematic illustrating separation into droplets (red) in the presence of a second component with fraction $\phi_S$. Blue is an additional separator to which red is compared for sensing the second component. Adapted from \cite{alston2026theoretical}. {\bf (d)} Dynamic sensory clusters. After budding from a yeast mother cell (blue asterisk), an a-type daughter cell's clustered sensory molecules (yellow arrowhead) move along the membrane toward an $\alpha$-cell mate. Adapted from \cite{wang2019mating}.}
\label{dynamic}
\end{figure*}

How does cell movement change our picture of cell sensing? The idea that sensing affects movement and movement affects sensing was implicit in Berg and Purcell's original analysis of bacterial chemotaxis \cite{berg1977physics}. This is because most bacteria perform temporal, not spatial, comparisons of concentration measurements \cite{block1982impulse, segall1986temporal}. The only way such a cell could sense a gradient is by moving along it ({\bf Fig.\ \ref{dynamic}a}, left). In fact, it has been argued that bacteria negotiate a tradeoff between efficient sensing and efficient movement, and the result ensures the highest minimum uptake of chemical across any concentration profile that they encounter \cite{celani2010bacterial}.

In contrast to bacteria, most eukaryotic cells perform spatial comparisons of concentration on different parts of their cell body \cite{parent1999cell}, which does not require cell movement for gradient sensing ({\bf Fig.\ \ref{dynamic}a}, right). Recent work has compared temporal and spatial sensing and revealed that the optimal chemotactic strategy is a combination of both \cite{rode2024information, alonso2024learning}, with spatial sensing more strongly weighted for larger, slower, and less persistent cells \cite{rode2024information}. Further, with a secreted coattractant in fluid flow, it has been found that temporal sensing promotes collective chemotaxis, whereas spatial sensing promotes individual chemotaxis \cite{gonzalez2023collective}.

\subsection{On-the-fly sensing}

Do cells integrate measurements for a prescribed time before acting? The alternative is that cells act as soon as they have sufficient data to do so, rather than waiting a prescribed time. This strategy has been termed on-the-fly decision-making \cite{siggia2013decisions} and builds upon optimal decision theory \cite{wald1945sequential}.

On-the-fly decision-making augments our picture of cell sensing. Previously we imagined a cell averaging its measurements over a fixed time $t$, after which it has obtained an estimation error ${\cal E}(t)$. On-the-fly sensing instead imagines that a cell only measures for long enough to rule out an alternative hypothesis. The time $t({\cal E})$ is then a function of the error threshold ${\cal E}$ with which the alternative is ruled out. {\bf Fig.\ \ref{dynamic}b} illustrates the conceptual difference, using concentration sensing as an example.

On-the-fly sensing has been applied to the problems of estimating a concentration from receptor binding events \cite{siggia2013decisions} and deciding between two cell fates based on a morphogen concentration in embryogenesis \cite{desponds2020mechanism}. The time required to reach a particular error threshold was found to be shorter than the time required to achieve an equivalent estimation error by fixed-time averaging, by factors of $2$$-$$3$ \cite{siggia2013decisions} and $10$ or more \cite{desponds2020mechanism}, respectively.

Cells' internal signaling networks are capable of both time-averaging with various weighting kernels \cite{govern2012fundamental} and on-the-fly decision-making \cite{siggia2013decisions, desponds2020mechanism}. A practical test of whether cells actually perform on-the-fly sensing would be to perturb the signal to make it easier to sense (e.g., increase a concentration) and determine whether cells wait for the same amount of time and act with smaller error, as would be expected for fixed-time averaging, or act after a smaller amount of time, as would be expected for on-the-fly sensing \cite{siggia2013decisions}.

Ultimately, on-the-fly sensing is a first-passage process: the cell responds at the first time an error threshold is met. First-passage processes occur in sensing and timing problems throughout cell biology \cite{iyer2016first, ghusinga2017first, gupta2018temporal}.

\subsection{Dynamic sensors}
How do cells dynamically adjust their sensors during sensing? A classic example comes from bacterial chemotaxis, where methylation causes receptors to adapt to the background concentration level, allowing cells to detect changes in concentration with high sensitivity \cite{tu2013quantitative}. Sensory adaptation is ubiquitous in cells, occurring both at the receptor level and in the signaling networks downstream \cite{ma2009defining, ferrell2016perfect, khammash2021perfect}.

Cells also adjust their receptors by redistributing them in space. An electric field is thought to redistribute receptors over the membrane as discussed in Section \ref{Selectric}, but this redistribution is passive, caused by electric mobility and diffusion rather than any direct action of the cell. In contrast, many cells actively redistribute their receptors, for example by endocytosis or clustering. Redistribution by endocytosis and diffusion has been shown to implement gradient sensing without the need for downstream signaling networks \cite{goetz2025emergent}.

Receptor clustering at regions of high membrane curvature can increase chemical intake \cite{wu2025optimal} and improve the precision of gradient sensing \cite{alonso2025local}. Amazingly, cells can turn clustering on and off with small changes in component concentrations by exploiting phase separation. This phenomenon, known as liquid-liquid phase separation or biomolecular condensate formation, has been the subject of intense research over the past two decades \cite{brangwynne2009germline, sengupta2007lipid, hyman2014liquid}. Recent work has investigated the precision of such a mechanism when used as a concentration sensor ({\bf Fig.\ \ref{dynamic}c}) and found that cells could distinguish concentration differences of 1\% within ten minutes \cite{alston2026theoretical}.

The combination of endocytosis and clustering can give rise to steerable sensory complexes. An important example occurs in yeast mating, where a cell finds a mate by detecting a secreted pheromone as discussed in Section \ref{Syeast}. The pheromone receptors form a cluster on the membrane, while endocytosis and recycling results in this cluster moving along the membrane toward the pheromone source ({\bf Fig.\ \ref{dynamic}d}). The cluster also contains the components necessary for the secretion of the degrading enzyme and the ultimate fusion of the two mating cells, leading it to be termed a gradient-tracking machine \cite{wang2019mating}. Because yeast cells are not motile, but rather they reorient and stretch to mate, these dynamic sensors play a dominant role in their gradient detection and response.

\section{Signal discrimination}
No cell can sense it all. Cells detect signals that improve their fitness and actively ignore others. How do they decide? Here we review recent progress on sensing in the presence of multiple signals.

\subsection{Selectivity}
\begin{figure*}
\includegraphics[width=.95\textwidth]{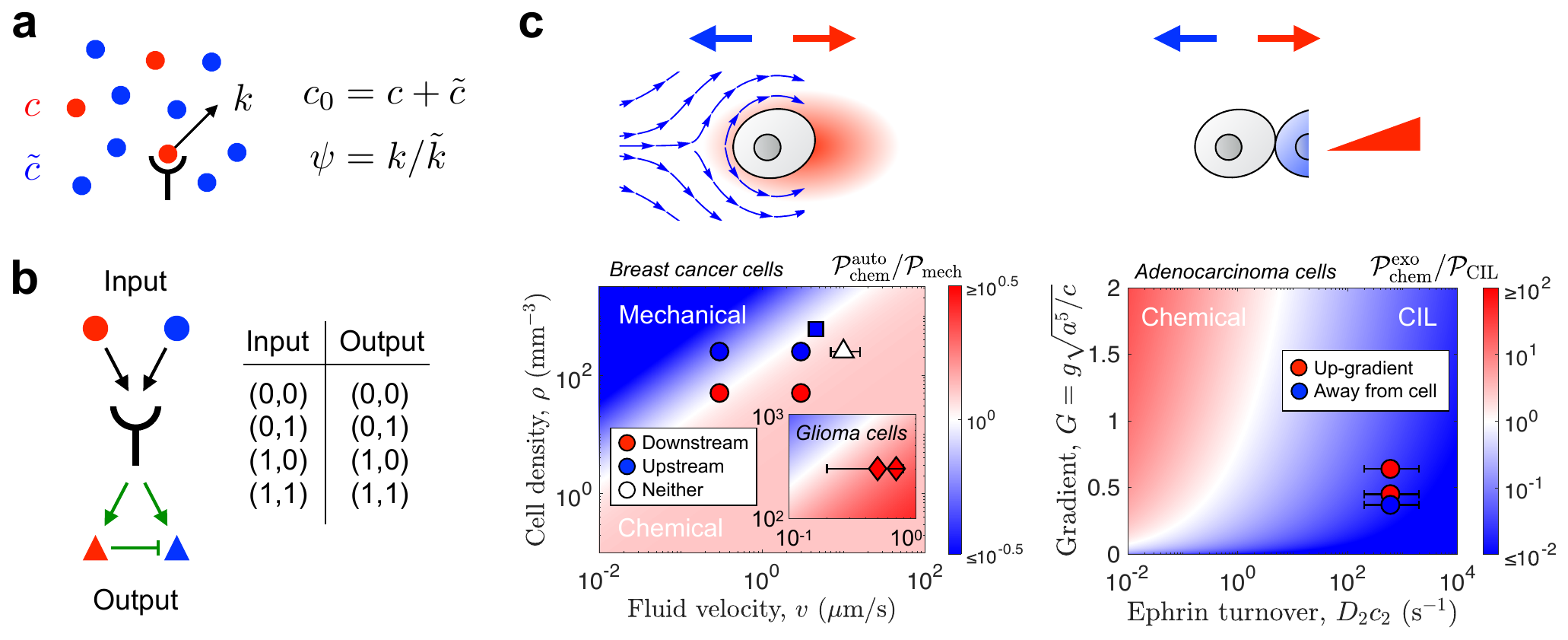}
\caption{Active sensing by signal discrimination. {\bf (a)} A receptor selects a concentration of interest $c$ within a total concentration $c_0$ using different unbinding rates \cite{mora2015physical}. {\bf (b)} A receptor multiplexes, encoding each of two input concentrations in each of two output concentrations, using a feed-forward network (green) \cite{de2011multiplexing}. {\bf (c)} Conflicting cues. Left: The ratio of Eq.\ \ref{flow} to Eq.\ \ref{autologous} (color map) predicts whether cells will go upstream (blue) or downstream (red), and experimental data are consistent. Right: The ratio of Eq.\ \ref{conc2} to Eq.\ \ref{grad} (color map) predicts whether cells will go up-gradient or away from another cell, and experimental data are not consistent, indicating a cell's strong internal preference for the gradient. Adapted from \cite{gonzalez2025quantifying}.}
\label{discrimination}
\end{figure*}

Concentration sensing occurs in the presence of chemicals of many types. Mora derived a physical limit for selecting a particular chemical of interest among many others \cite{mora2015physical} ({\bf Fig.\ \ref{discrimination}a}). For a receptor of size $a$ sensing a concentration of interest $c$ within a total concentration $c_0$, the estimation error scales as
\begin{equation}
\label{conc_disc}
\frac{\delta c}{c} \sim \sqrt{\frac{(c_0/c)^\lambda}{caDt}},
\end{equation}
where $\lambda = \psi/(1-\psi)$ when $\psi < 1/2$, and $\lambda = 1$ when $\psi > 1/2$. Here, $\psi < 1$ is the ratio of the rate at which the receptor unbinds the chemical of interest vs.\ the other chemicals; a more specific receptor releases the chemical of interest at a lower rate, lowering $\psi$.

Comparing to Eq.\ \ref{conc}, we see that the presence of other chemicals increases the estimation error via the factor $(c_0/c)^\lambda\ge1$. As expected, this factor goes to one when there is only the chemical of interest ($c_0 \to c$) or the receptor's specificity to this chemical is infinite ($\psi \to 0$). Since selecting a chemical of interest over others can be framed as a decision, this process has also been analyzed with on-the-fly sensing as introduced above \cite{siggia2013decisions, mora2015physical}. Interestingly, Mora showed that the selection can be performed mechanistically by a class of intracellular signaling networks known as kinetic proofreading \cite{mora2015physical}. Kinetic proofreading requires energy dissipation \cite{hopfield1974kinetic}, underscoring the active nature of selective concentration sensing.

\subsection{Multiplexing}
Can a cell sense multiple signals with the same component simultaneously? This strategy, termed multiplexing \cite{de2011multiplexing}, has been shown to be possible with a single receptor. If the signals are encoded in the receptor binding time series, the decoding can be done using a kinetic proofreading network \cite{singh2017simple}, similar to the case of selecting one signal as discussed above. If the signals are encoded in the component concentrations, the decoding can be done using a feed-forward network \cite{de2011multiplexing} ({\bf Fig.\ \ref{discrimination}b}). A key limitation on multi-signal sensing is network saturation, since only so much information can pass through a given intracellular component per time \cite{moon2021signal}. Saturation can lead to surprising effects, such as a cell responding to two gradient signals when presented individually but not when presented together in the same direction \cite{moon2021signal, saha2022deduction}.

\subsection{Conflicting cues}
When a cell is presented with conflicting sensory cues, to what extent do the physical limits discussed herein predict its response? This question was recently addressed for cancer cells subjected to pairs of conflicting cues \cite{gonzalez2025quantifying}. One pair arises during fluid flow sensing, as discussed in Sections \ref{Sflow} and \ref{Sautologous}: mechanical sensing of the flow promotes upstream migration whereas autologous chemotaxis promotes downstream migration \cite{polacheck2011interstitial, polacheck2014mechanotransduction}. Taking the ratio of Eq.\ \ref{flow} to Eq.\ \ref{autologous} and inserting measured parameter values produces a phase diagram of migration direction as predicted by sensory precision alone, and cells' migration behavior was found to be consistent with this prediction \cite{gonzalez2025quantifying} ({\bf Fig.\ \ref{discrimination}c}, left).

However, another pair of conflicting cues arises when a concentration gradient promotes migration in one direction whereas collision with another cell promotes migration in the other direction in a process called contact inhibition of locomotion (CIL) \cite{lin2015interplay}. CIL is triggered by one cell sensing the concentration $c_2$ of a chemical on the membrane of the other cell \cite{lin2015interplay, wang2024limits}. In the membrane-to-membrane contact region, this concentration sensing problem is essentially two-dimensional. Therefore, the argument leading to Eq.\ \ref{conc} must be changed to read $n\sim(\delta n)^2\sim c_2a^2$, where $a^2$ is now the area of the contact region \cite{gonzalez2025quantifying}. Eq.\ \ref{conc} then becomes
\begin{equation}
\label{conc2}
\frac{\delta c_2}{c_2} \sim \frac{1}{\sqrt{c_2D_2t}},
\end{equation}
where $D_2$ is the diffusion coefficient on the membrane.

Taking the ratio of Eq.\ \ref{conc2} to Eq.\ \ref{grad} and inserting measured parameter values again produces a phase diagram of migration direction, but here cells' migratory behavior was seen to violate the phase boundary by several orders of magnitude \cite{gonzalez2025quantifying} ({\bf Fig.\ \ref{discrimination}c}, right). Thus, here cells respond to a signal (the gradient) even when it is sensed with a precision that is far less than a conflicting signal. This behavior reflects an internal preference and in this particular case can be rationalized from the structure of the cells' internal signaling network \cite{gonzalez2025quantifying}. We see that while physical limits set the maximum information that can be sensed, a cell's internal composition determines whether this information directly shapes the cell's response, or whether instead some signals are deprioritized in favor of others. 

\section{Future directions}
We have reviewed recent progress on the physics of active cell sensing. As summarized in Fig.\ \ref{overview}, we have seen that some active processes, such as cell-cell coordination (Eqs.\ \ref{concN} and \ref{gradN}), environmental modification (Eqs.\ \ref{degrade}-\ref{autologous2}), and dynamic updating (Fig.\ \ref{dynamic}b), allow cells to acquire more sensory information than passive sensors. On the other hand, we have seen that signal discrimination allows cells to either enhance or suppress a signal's information depending on the context (Fig.\ \ref{discrimination}c). Where might this field go next?

Recent work has measured the sensory information acquired by swimming bacteria and shown that it is orders of magnitude less than the physical limit implied by Eq.\ \ref{conc} \cite{mattingly2021escherichia, mattingly2026coli}. Thus, the argument of Berg and Purcell from fifty years prior \cite{berg1977physics}, that a bacterium knows definitively before changing direction whether the concentration has increased or decreased, appears incorrect. Instead, bacteria appear to require many such directional changes before acquiring sufficient information, which still results in a statistical drift toward higher concentrations, but more slowly \cite{mattingly2026coli}. Reasons for such poor information acquisition include that competing fitness goals may take priority \cite{mattingly2026coli} or that poor information need not always correspond to poor migratory performance \cite{endres2026robust}. What will careful measurement of acquired information reveal in other organisms? Are eukaryotic cells, which sense chemical gradients spatially rather than temporally, similarly far from the physical limit?

Active sensing requires energy. The energy requirements for reaching the physical limits to sensing have been explored extensively in the contexts of biochemical networks \cite{govern2014optimal, govern2014energy}, adaptation \cite{lan2012energy}, intracellular signaling \cite{bryant2023physical} and more. Less explored are the energy requirements for extending the limits themselves, using some of the strategies reviewed here. How much energy does it take to secrete a reporter molecule or a degrading enzyme, or to communicate sensory information to other cells? Is energy better spent internally to approach a sensing limit, or externally to extend it?

Sensing is critical for other biological behaviors. To what extent do sensory limits shape or constrain these behaviors? For example, an area of accelerating interest is understanding the computational capacity of a cell's internal signaling network. One way to characterize this capacity is by how well the network can be trained by inputs \cite{trifonova2025trainable} or how many inputs it can classify \cite{floyd2025limits}. To what extent does a cell's ability to sense these inputs constrain these measures of computational capacity? Other examples occur in active matter. A collective phenomenon known as motility-induced phase separation \cite{cates2015motility} changes dramatically when the motile agents sense concentration gradients \cite{zhao2023chemotactic}. To what extent are other active matter phenomena altered or limited by sensory capabilities \cite{ziepke2022multi}?

Sensing is a cell's window to the world. It remains to be seen how broadly the physics of sensing shapes the biology of the cell.

\acknowledgments
This work was supported by National Science Foundation grant numbers PHY-2118561 and MCB-2341919.


%

\end{document}